# Magnetocaloric Effect and Magnetic refrigeration: analytic and numeric study

A. Boubekri, M.Y. El Hafidi and M. El Hafidi*
Laboratory of Condensed Matter Physics
Hassan II University of Casablanca
B.P 7955, Casablanca, Morocco
*e-mail : mohamed.elhafidi@univh2c.ma

**Abstract**

This work aims to present an analytical and numerical study of the magnetocaloric effects (MCE) providing realistic proposals about materials that should be chosen in the design of new refrigerator appliances around the room temperature.

Starting from a spin Hamiltonian including the exchange interaction, the single-ion anisotropy and the applied magnetic field terms, we have calculated the partition function at a given temperature and derived a set of relevant physical quantities as magnetization, magnetic entropy and specific heat and analysed their behaviour with atomic parameters as spin, exchange and anisotropy. Using numerical programs that we developed by ourselves, we were able to better elucidate the role of each microscopic parameter in order to reinforce the relative cooling power (RCP) and give rise to optimal performances of the refrigerant compound. This approach could be extended to composite materials underscoring a giant MCE at room temperature.

I.  **Introduction**

Efforts devoted to developing energy-efficient technologies, especially in refrigeration around the room temperature, have given the magnetocaloric effect (MCE) a hopeful interest for a large scientific community.  The number of scientific publications on this topic is increasing exponentially in the recent years, proving the high outlooks related to this field of research [1]. Thanks to their increased energy efficiency, magnetic refrigerators are expected to have more reduced environmental fallout comparatively with those based on the gas compression-expansion cycles, as they do not involve ozone-depleting or greenhouse effect related gases. Also, the lack of a large compressor in the magnetic refrigerator allows less vibration and noise [2].

The MCE is associated to a significant change in magnetization close to the working temperature of the refrigerant material. Previously, magnetic refrigeration at low temperatures relied on paramagnetic salts, as their magnetization increases remarkably at very low temperature [3]. However, for applications at temperatures close to room temperature, a different approach had to be found: the existence of a phase transition in the material close to the working temperature would produce the required abrupt change in magnetization. From the physical point of view, magnetic refrigerant materials can be classified by the type of phase transition that they undergo.

It can be a second order magnetic phase transition, SOPT (like the ferro-paramagnetic transition of a ferromagnetic material at its Curie temperature), which is characterized by the lack of thermal and magnetic hysteresis, and in which the magnetization decreases continuously to zero. Pure gadolinium is a paradigmatic example of a magnetic refrigerant undergoing a phase transition of this kind. But phase transitions can also be of the first order type, FOPT, in which magnetization shows a rude change at the transition temperature, usually associated to a magneto-structural phase transition, giving rise to the giant magnetocaloric effect (GMCE), with $Gd_5Si_2Ge_2$ being the typical case of this kind of magnetic refrigerant materials [4]. However, although the large abrupt change in magnetization causes a correspondingly giant magnetic entropy change, this appears at the cost of thermal and magnetic hysteresis, which should be avoided in order to be able to apply these materials in refrigerator appliances.

This work aims to present an analytical and numerical study of the MCE providing rational proposals concerning materials that should be chosen in the design of new refrigerator appliances around the room temperature. Our numerical computations and simulations show that the proposed approach enhances considerably the MCE and gives rise to optimal performances.

The paper is organized as follows: In Sec. II, we introduce the spin Hamiltonian for a ferromagnetic system with exchange and single-ion anisotropy interactions. We explain how to obtain physical relevant parameters using the Mean Field Approximation (MFA), noting that we can consider a multiphase system with more than one kind of spin. The dependence of the magnetic entropy on the magnetic ordered structures is shown. In addition, we discuss the relation between the magnetic entropy, the specific heat and the magnetization behaviours. In Sec. III, the magnetic refrigeration efficiency in the isothermal and adiabatic magnetization processes is considered. To assess the efficiency, we calculate the isothermal magnetic entropy change and the adiabatic temperature change. In order to investigate the efficiency in the isothermal demagnetization process, we introduce a characteristic parameter called Relative Cooling Power (RCP) giving a direct information about the refrigerant performances. In Section IV, results are presented and discussed. The Section V is devoted to a general conclusion & perspectives.

II. **The model**

First, we consider a ferromagnetic single-system which is described by the following spin Hamiltonian:

$$H_\alpha = -\sum_{<i,j>} J_{\alpha,ij} \vec{S}_{i,\alpha} \cdot \vec{S}_{j,\alpha} - D_\alpha \sum_i {S_{i,\alpha}^z}^2 - g\mu_B B_0 \sum_i S_{i,\alpha}^z \quad \text{(1)}$$

where $J_{\alpha,ij}$ is the exchange coupling constant among nearest-neighbours ($J_\alpha>0$), $D_\alpha$ is the single-ion anisotropy constant for a site ($\alpha$,i), $\vec{S}_{i,\alpha}$ (respectively $\vec{S}_{j,\alpha}$) are the spin operators acting on the site i (respectively j) and $B_0$ is the applied magnetic field along the z-axis. Here, g is the Landé factor, $\mu_B$ is the Bohr magnetron and $B = \mu_0 H_0$ where $\mu_0$ is the magnetic permeability. The first double summation runs over all pairs of nearest neighbours.

To make progress while solving eqn. (1), it is necessary to make some approximations. We define an effective molecular field action on the $i^{th}$ site as follows [5]:

$$H_{MF} = \frac{1}{g\mu_B} \sum_{j \in nn} <S_{\alpha,j}> J_{\alpha,ij} \qquad (2)$$

Therefore, the exchange interaction is replaced by the effective molecular field $H_{MF}$ produced by the neighbouring spins. We are now able to treat this problem as if the system was a simple paramagnet located in a magnetic field $H_0 + H_{MF}$. The effective Hamiltonian may be rewritten as:

$$H_\alpha = -g\mu_0\mu_B \sum_i S^z_{i,a} (H_0 + H_{MF}) - D_\alpha \sum_i S^{z^2}_{i,\alpha} \qquad (3)$$

The assumption supporting this approach is that all magnetic ions apply the same molecular field. However, this may be rather questionable, particularly at temperatures close to a magnetic phase transition. For a ferromagnet, the molecular field will act so as to align neighbouring magnetic moments. Since the molecular field measures the effect of the system ordering, we can assume that $H_{MF} = \lambda M$, where:

$$\lambda = \frac{zJ}{(g\mu_B)^2 n} \qquad (4)$$

Thus, $\lambda$ is a parameter which describes the strength of the molecular field as a function of the magnetization (for a ferromagnet, $\lambda > 0$). Here, z and n denote respectively the coordination number and the volume spin concentration which are together related to the intrinsic crystallographic structure of the material.

Within this assumption, our Hamiltonian may be similarly rewritten as:

$$H_\alpha = \sum_i H_{\alpha,i} \qquad (5)$$

where the single-ion Hamiltonian $H_{\alpha,i} = -g\mu_B S_{i,\alpha}(B_0 + B_{MF}) - DS_i^2$ is used with corresponding eigenvalues $E_i = -g\mu_B m_i(B_0 + B_{MF}) - Dm_i^2$ where $-S \leq m_i \leq S$.

In addition, it may be observed that all individual Hamiltonians $H_{i,\alpha}$ commute mutually permitting us to write the partition function of the whole system $\alpha$:

$$Z_\alpha = \prod_i Z_{i,\alpha} \qquad (6)$$

where $Z_{i,\alpha}$ is partition function of the particle i and the canonical sum is performed over all the stationary states of $H_{i,\alpha}$

$$Z_{i,\alpha} = \sum_{\{m_i\}} e^{\beta g\mu_B m_i \mu_0(H_0 + H_{MF}) - \beta D m_i^2} = \sum_{m_i=-S}^{m_i=+S} e^{y_1 m_i + y_2 m_i^2} \qquad (7)$$

where we have take $y_1 = \beta g\mu_B(B_0 + B_{MF}) = \beta g\mu_B(B_0 + \lambda M); y_2 = \beta D; \beta = 1/k_B T$, $k_B$ being the Boltzmann constant.

A simple calculation allows us to write $Z_{i,\alpha} = F_S(y_1, y_2)$ where $F_S(y_1, y_2)$ is a spin-dependent function :

$F_{\frac{1}{2}}(y_1, y_2) = 2e^{\frac{y_2}{4}} \cosh(\frac{y_1}{2})$ for S= ½

$F_1(y_1, y_2) = 2e^{y_2} \cosh(y_1) + 1$ for S=1

$F_{\frac{3}{2}}(y_1, y_2) = 2e^{\frac{9y_2}{4}} \cosh(\frac{3y_1}{2}) + 2e^{\frac{y_2}{4}} \cosh(\frac{y_1}{2})$ for S=3/2

$F_2(y_1, y_2) = 2e^{4y_2} \cosh(2y_1) + 2e^{y_2} \cosh(y_1) + 1$ for S=2

and so on.

Once the calculation of $Z_\alpha$ is carried out, one can derive the Helmholtz free energy for n spins per unit volume using the expression $F_\alpha = -n k_B T \ln(Z_\alpha)$ and deduce the other relevant physical quantities. Thus, the magnetization, the magnetic entropy per unit volume and the magnetic specific heat will be expressed respectively by:

$$M_\alpha(B,T) = -(\frac{\partial F_\alpha}{\partial B})_{T,V} = nk_B T (\frac{\partial \ln(Z_\alpha)}{\partial B})_{T,V} \quad (8)$$

and

$$S_M = -(\frac{\partial F_\alpha}{\partial T})_{B,V} = n\frac{\partial (k_B T \ln(Z_\alpha))}{\partial T})_{B,V} \quad (9)$$

$$C_M = T(\frac{\partial S_M}{\partial T})_M \quad (10)$$

By performing calculation for several given spin values, one finds detailed expressions of these thermodynamically variables as follows:

for magnetization ($m = M/M_s$)

$$m_{1/2} = \frac{1}{2} \tanh\left(\frac{y_1}{2}\right) \quad \text{for S=1/2} \quad (11a)$$

$$m_1 = \frac{2\sinh(y_1)}{2\cosh y_1 + \exp(-y_2)} \quad \text{for S=1} \quad (11b)$$

$$m_{3/2} = \frac{1}{2} \frac{3\sinh\left(\frac{3y_1}{2}\right) + \exp(-2y_2)\sinh\left(\frac{y_1}{2}\right)}{\cosh\left(\frac{3y_1}{2}\right) + \exp(-2y_2)\cosh\left(\frac{y_1}{2}\right)} \quad \text{for S=3/2} \quad (11c)$$

for entropy

$$S_{1/2} = k_B \ln\left[2\cosh\left(\frac{y_1}{2}\right)\right] - \left(\frac{y_1}{2\beta T}\right)\tanh\left(\frac{y_1}{2}\right)$$

$$S_1 = k_B \ln[1 + 2\exp(y_2)\cosh(y_1)] + \frac{-2D\exp(y_2)\cosh(y_1) - 2\exp(y_1)\left(\frac{y_1}{\beta}\right)\sinh(y_1)}{T(1 + 2\exp(y_2)\cosh(y_1))}$$

**(12)**

and for heat capacity:

$$C_{1/2} = \left(\frac{y_1}{2\beta T}\right)^2 \frac{1}{k_B \cosh^2\left(\frac{y_1}{2}\right)}$$

$$C_1 = 2\exp(y_2)\left[\frac{(y_2 - y_1)\cosh(y_1)}{k_B(1 + 2\exp(y_2)\cosh(y_1))} + \left(\frac{D\cosh(y_1) + \left(\frac{y_1}{\beta}\right)\sinh(y_1)}{1 + 2\exp(y_2)\cosh(y_1)}\right)^2\right] \quad \text{(13)}$$

However, for magnetic entropy, we must be very careful. In fact, the total entropy of a magnetic material is the sum of the contributions of magnetic moments $S_{mag}$ the vibrational crystal lattice part $S_{lat}$ and the conduction electrons part $S_{el}$. Thus, global entropy is written at a constant pressure:

$$S(T,H) = S_{mag}(T,H) + S_{lat}(T) + S_{el}(T) \quad (14)$$

It is assumed that only the magnetic entropy is dependent on the magnetic field. Besides, the entropy of the crystal lattice is giving by the Debye model [6]:

$$S_{lat}(T) = Nk_B\left[-3\ln\left(1 - \exp\left(\frac{T_D}{T}\right)\right) + 12\left(\frac{T_D}{T}\right)^3 \int_0^{\frac{T_D}{T}} \frac{x^3}{\exp(x) - 1}dx\right] \quad (15)$$

where $T_D$ is the Debye temperature, while the entropy of electrons is usually described by the Somerfield model [7]:

$$S_{el}(T) = \frac{\gamma T}{M} \quad (16)$$

where M and $\gamma$ denotes respectively the molar mass and Somerfield constant. Thus, for a rigorous analysis of experimental data, these contributions must be highlighted and carefully controlled.

Finally, note that the magnetization M is present implicitly therein eqs. (8, 9 and 10). For solving this set of self-equations numerically, we have performed a computational program permitting us to simulate the evolution of these pertinent quantities with both magnetic field and temperature for a given set of microscopic parameters as exchange interaction, spin magnitude or coordination of magnetic sites.

### III. Thermodynamic study and Refrigeration parameters

To understand the physical origin of the magnetocaloric effect, it is useful to recall the basically thermodynamic properties of a magnetic material immersed in a magnetic field. The thermodynamic potential adapted to the description of such a system is the free enthalpy G which is expressed in terms of the internal energy U, extensive variables: the magnetic entropy $S_M$, magnetization M and intensive variables: temperature T and magnetic induction B which is directly related to the external magnetic field $H_0$ by $B = \mu_0 H_0$ [8].

For a solid system, we can neglect any effect due to the volume and pressure and we can write the exact total differential of the free energy as follows:

$$dG = \left(\frac{\partial G}{\partial T}\right) dT + \left(\frac{\partial G}{\partial B}\right) dB = -MdB - S_M dT \quad (17)$$

Cross partial second derivatives of an exact differential being identical, hence one obtains the Maxwell-Weiss relation [6]:

$$\left(\frac{\partial M}{\partial T}\right)_B = \left(\frac{\partial S_M}{\partial B}\right)_T \quad (18)$$

which after integration gives the magnetic entropy change for an isothermal demagnetization process, when the magnetic field passes from $B_0$ to $B_1$ at a given temperature as:

$$\Delta S_M = \int_{B_0}^{B_1} \left(\frac{\partial M}{\partial T}\right)_B dB \quad (19)$$

Besides, using equation (14) and the fundamental equations (when the magnetic entropy decreases the lattice heat increases):

$$C_M dT = \delta Q_M \text{ and } \delta Q_M = -T dS_M \quad (20)$$

we get for infinitesimal adiabatic temperature rise:

$$dT = -\left(\frac{T}{C_M(T,B)}\right)_M \left(\frac{\partial M}{\partial T}\right) dB \quad (21)$$

After integrating Eq. (21) for an adiabatic transformation, the magnetocaloric effect (MCE) can be expressed by [7]:

$$MCE = \Delta T = -\int \left(\frac{T}{C_M(T,B)}\right)_M \left(\frac{\partial M}{\partial T}\right)_B dB \quad (22)$$

where $\left(\frac{\partial M}{\partial T}\right)_B$ is the thermal magnetization for a fixed magnetic field. It is worth recalling that the temperature T and magnetic field $B_0$ dependences of the magnetic entropy $S_M(T, B_0)$ are

---

crucial for designing magnetic refrigeration cycles. So, producing a magnetocaloric material that possesses a large magnetic entropy change ($\Delta S_M$) over a wide temperature range ($\Delta T$), i.e., a large refrigerant capacity, is highly required for magnetic refrigeration applications.

The relative cooling power (RCP) has been often used as a standard for good magnetic refrigeration materials [1, 9]:

$$\text{RCP}(B_0 \rightarrow B_1) = \Delta S_{M\,max}(B_0 \rightarrow B_1) \times \Delta T_{1/2}(B_0 \rightarrow B_1) \quad (23)$$

where $\Delta S_{M\,max}(B_0 \rightarrow B_1)$ and $\Delta T_{1/2}(B_0 \rightarrow B_1)$ are the maximum value and the full width at half maximum of $\Delta S_M(B_0 \rightarrow B_1)$ at given $B_0$ and $B_1$, respectively.

### IV. Results and discussion

Temperature dependence of the spontaneous magnetization for different values of spin S is shown in Fig.1. For simplicity, we have plotted the relative magnetization $M/M_s$ versus the reduced temperature $T/T_c$.

The shape of the curves is slightly different yet a general trend is present:
for $T > T_c$ the magnetization vanishes $M=0$, for $T < T_c$: $M>0$ the magnetization values are quite stable at low temperatures and at $T = T_c$, the magnetization is not continuously differentiable. Thus, the phase transition is of second order. Also, it is worth to notice that for a given spin system, $T_c$ shifts to higher temperatures for stronger exchange interaction J. In addition, $T_c$ is enhanced for higher spin systems.

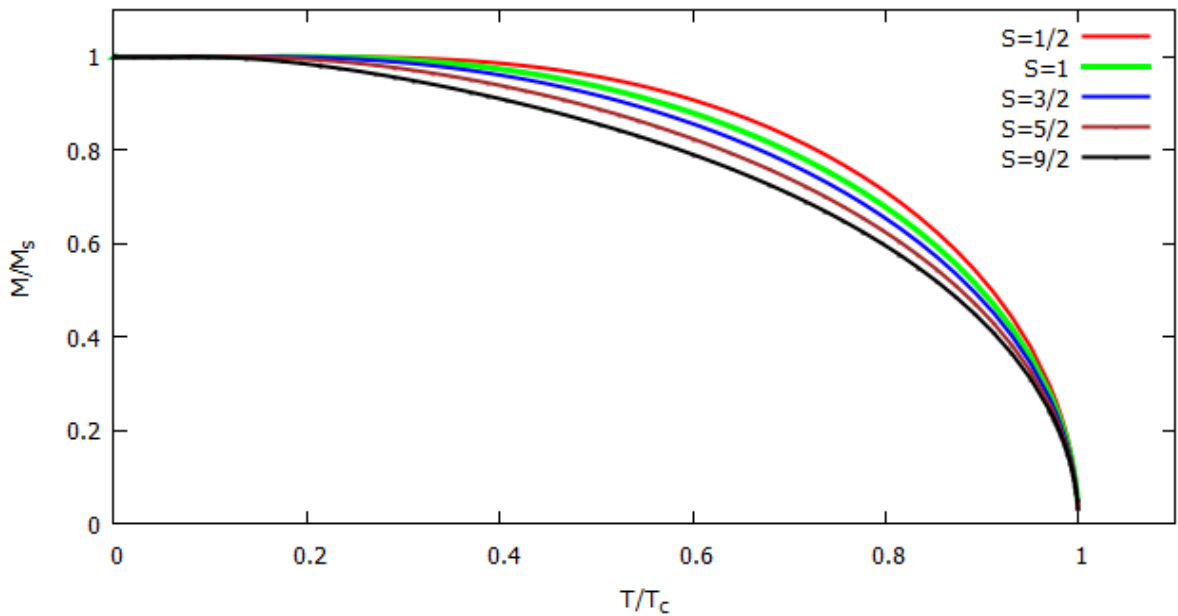

*Fig.1 : Relative magnetization as a function of the reduced temperature $T/T_c$ for different values of spin S.*

Temperature dependence of the relative magnetization $M/M_S$ for a given spin system under an applied magnetic field expressed by the reduced parameter $b = g\mu_B BS/J = 0, 0.5, 1, 1.5$ and $2$ is displayed in Fig.2. Since the considered system is ferromagnetic, a non-vanishing magnetization persists even above the critical temperature when an external magnetic field is applied, which means that different spins remain relatively correlated one to the other and the transition is pushed back to high temperatures.

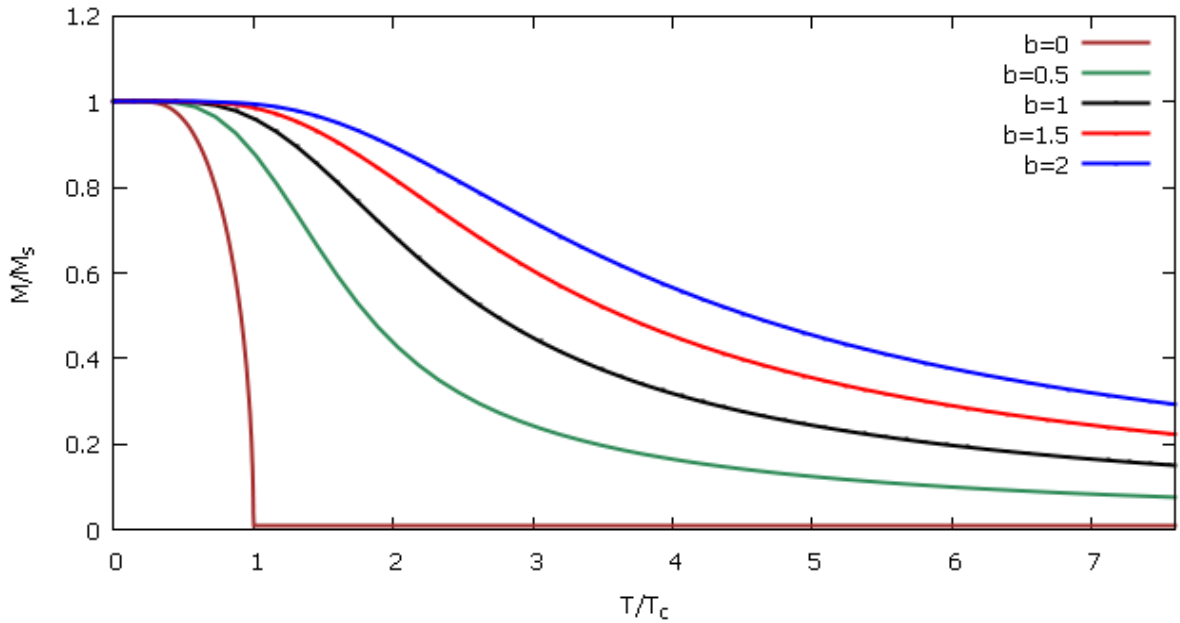

*Fig.2 : Relative magnetization as a function of the reduced temperature $T/T_c$ for different values of magnetic field (*$b = g\mu_B BS/J$*) for a given spin S . A vanishing magnetization only occurs for b=0. For increasing field, the transition is delayed to high temperature.*

The magnetic entropy curves versus temperature under different external magnetic fields for a given spin system is reported in Fig. 3.
While applying a magnetic field, magnetic entropy takes time before attaining its limit value.

**Magnetocaloric Effect and Magnetic refrigeration: analytic and numeric study**
A. Boubekri, M.Y. El Hafidi and M. El Hafidi

*3rd International Congress On Advanced Technologies'2017- April 12 - 14, 2017, Safi – Morocco*

---

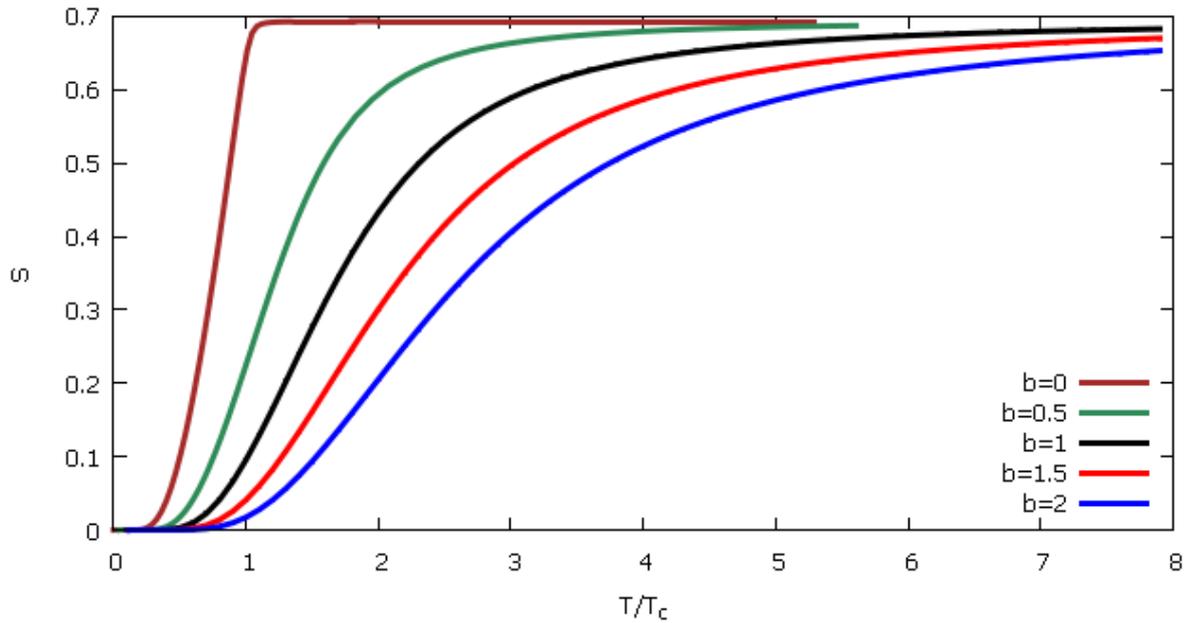

*Fig. 3:* Magnetic entropy curves versus temperature under different external magnetic fields.

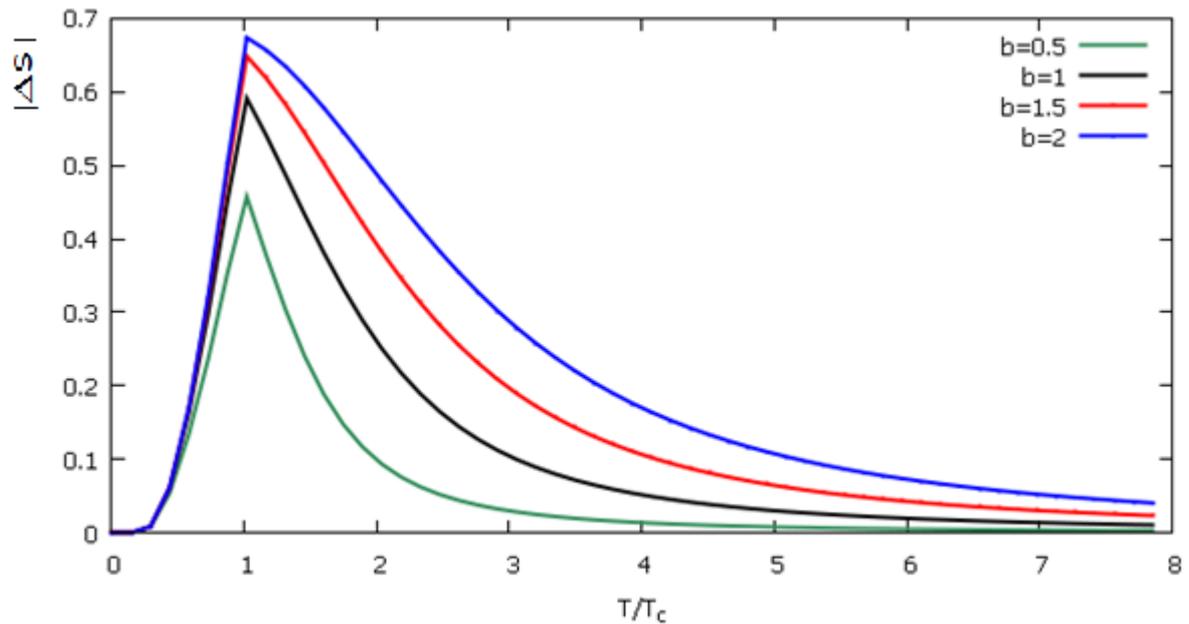

*Fig.4 : Absolute value of* $|\Delta S_M|$ *as a function of temperature for different values of the applied magnetic field for a system with S=1/2.*

The entropy changes as function of temperature, derived from magnetic isotherms through the Maxwell relations (18) are displayed in Fig. 4 for S=1/2. The absolute value $|\Delta S_M|$ reaches a maximum around $T_C$, under various magnetic fields. This maximum increases strongly with the applied magnetic field. $|\Delta S_M|$ is also displayed for different values of the system spin. As shown in Fig. 5, $|\Delta S_M|$ grows significantly with the spin size. Actually, the value of $\Delta S_M$ is negative in the entire temperature range and is extended over a wide range of temperature around the Curie temperature, which is useful for a room temperature or above room temperature magnetic refrigeration process [10].

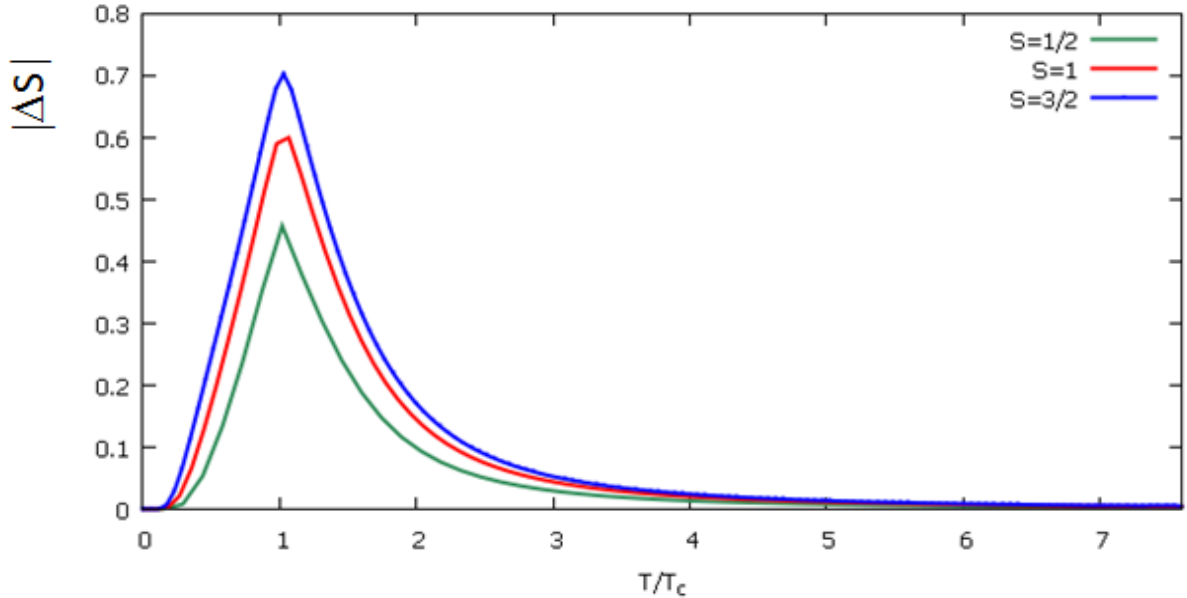

Fig. 5: Temperature dependence of the magnetic entropy change for different values of spin.

Recalling that the temperature T and magnetic field $B_0$ dependences of the magnetic entropy $S_M(T,B)$ are important for designing magnetic refrigeration cycles. So, producing a magnetocaloric material that possesses a large magnetic entropy change $|\Delta S_M|$ over a wide temperature range ($\Delta T$), i.e., a large refrigerant capacitY, is highly required for magnetic refrigeration applications [11]. Consequently, both large isothermal entropy and large adiabatic temperature changes are required for good magnetic refrigeration materials.

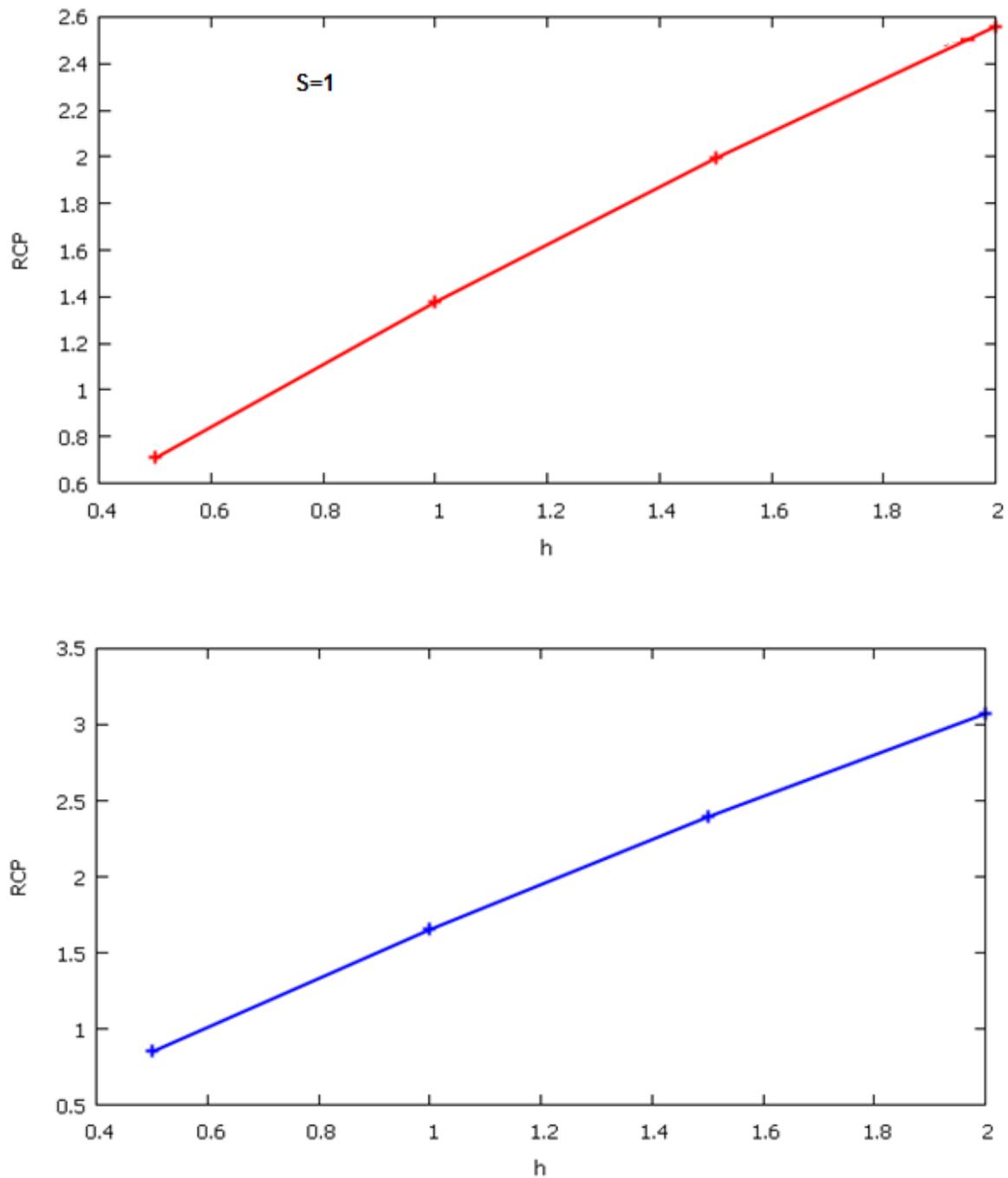

Fig. 6(a): the field dependence of RCP of a spin-1 compound
Fig. 6(b): the field dependence of RCP of a spin-3/2 compound

The larger RCP values of a given material may result from the increase of both $|\Delta S_M|$ and $\Delta T_{1/2}$.

Finally, we consider a multi-phase material with two kinds of spin in order to enhance the RCP, i.e. improving $|\Delta S_M|$ and $\Delta T_{1/2}$. This protocol would be able to describe the magnetic refrigeration efficiency of two kinds of ferromagnets as in composite materials [12].

The corresponding Hamiltonian rewrites as $H = \sum_\alpha H_\alpha$ where $[H_\alpha, H_{\alpha'}] = 0$, the partition function becomes then $Z(T,B) = \prod_\alpha Z_\alpha(T,B)$. Consequently, one can add the extensive variables as $\Delta S_{M\alpha}$. A simple composite material that we have considered is viewed as two superposed mediums with spins $S_1$ and $S_2$ and respective concentrations c and (1-c). The obtained results for a composite material with spin ½ and 1 confirm this conjecture.

The large temperature span and RCP values previewed with mixed spin systems in this work make them suitable for potential use in the efficient Ericsson-cycle magnetic refrigeration and motivates the experimental investigations in this way.

Possible candidates for such systems are the multilayer composites of $Fe_{88-x}Nd_xCr_8B_4$ alloys with various Nd substitutions for Fe (x=0.05, 0.0 8, 0.10, 0.12 and 0.15) [13].

Magnetic polymer composite materials have raised the attention of the scientific community in the last decades mainly because of their biomedical applications. These materials, synthesized by embedding magnetic particles into a polymer matrix, have light weight and high shape-flexibility and are commonly used for bio-magnetic separations processes.

V. **Conclusion & perspectives**

The magnetocaloric effect in ferromagnetic systems is studied using the Mean Field approximation and by performing a numerical simulation. The maximum of the magnetocaloric effect occurs around the Curie temperature $T_C$. Spin, exchange interaction and magnetic field work all together to enhance the MCE and RCP that are the most pertinent parameters for refrigeration. This study confirms also, that when a mixture spin system is elaborated with two kinds of spin, as in the composite compounds, the RCP is strongly enlarged.